\documentclass[aps,showpacs,twocolumn,floatfix,superscriptaddress,nofootinbib]{revtex4}
\usepackage[dvips]{graphicx}
\usepackage{amsmath}

\begin{document}


\preprint{KIAS-P05034}
\preprint{YITP-05-25}
\preprint{VPI-IPPAP-05-04}
\preprint{hep/ph-0510082}

\title{Leptonic CP Violation Search and the Ambiguity of $\delta m^{2}_{31}$}

\author{Masafumi~Koike}
\email{koike@vt.edu}
\affiliation{%
Institute for Particle Physics and Astrophysics,
Physics Department, Virginia Tech, Blacksburg VA 24061, USA}
\author{Naotoshi~Okamura}
\email{okamura@yukawa.kyoto-u.ac.jp}
\affiliation{%
Yukawa Institute for Theoretical Physics,
Kyoto University, Kyoto 606-8502, Japan}
\affiliation{%
Korea Institute for Advanced Study,
207-43 Cheongnyangi 2-dong, Dongdaemun-gu, Seoul 130-722, Korea}
\author{Masako~Saito}
\affiliation{%
Institute for Particle Physics and Astrophysics,
Physics Department, Virginia Tech, Blacksburg VA 24061, USA}
\author{Tatsu~Takeuchi}
\email{takeuchi@vt.edu}
\affiliation{%
Institute for Particle Physics and Astrophysics,
Physics Department, Virginia Tech, Blacksburg VA 24061, USA}

\date{\today}

%
%
\begin{abstract}
\noindent
We consider a search for the CP-violating angle $\delta_{\textrm{CP}}$
in long baseline neutrino oscillation experiments.
We show that the subleading $\delta_{\textrm{CP}}$-dependent terms in
the $\nu_{\mu} \to \nu_{\textrm{e}}$ oscillation probability can be
easily obscured by the ambiguity of the leading term which depends on
$| \delta m^{2}_{31} |$.
It is thus necessary to determine the value of $\delta m^{2}_{31}$
with a sufficient accuracy.
The $\nu_{\mu}$ survival events, which can be accumulated
simultaneously with the $\nu_{\textrm{e}}$ appearance events, can
serve for this purpose owing to its large statistics.
Therefore, the combined analysis of $\nu_{\textrm{e}}$ appearance and
$\nu_{\mu}$ survival events is crucial to provide a restrictive
constraint on $\delta_{\textrm{CP}}$.
Taking a test experimental setup, we demonstrate in the
$\delta_{\textrm{CP}}$-$\delta m^{2}_{31}$ plane that
the analysis of $\nu_{\textrm{e}}$ appearance events leads to less
restrictive constraints on the value of $\delta_{\textrm{CP}}$ due to
the ambiguity of $\delta m^{2}_{31}$
and that
the combined analysis efficiently improves the constraints.
\end{abstract}

\maketitle

%
%

\section{Introduction}
\label{sec:intro}

Neutrino oscillation has been strongly suggested by an accumulating
number of experiments using a variety of neutrino sources~%
\cite{bib:solar,bib:atmospheric,bib:terrestrial}, providing rich
information on the flavor structure of the lepton sector.
The mass- and the mixing-parameters of neutrinos are nonetheless still
not completely known;
following the definitions of Ref.~\cite{Eidelman:2004wy},
the unknown parameters include
the value of the mixing angle $\theta_{13}$,
the CP-violating angle $\delta_{\textrm{CP}}$,
and the sign of $\delta m^{2}_{31}$.
The current upper bound on $\theta_{13}$~\cite{Apollonio:2002gd}
is expected to be improved
down to $\sin^{2} 2 \theta_{13} \sim 0.01$
by the next generation of nuclear 
reactor~\cite{Anderson:2004pk,Huber:2004ug,bib:reactor-minakata,Huber:2003pm}
and accelerator experiments~\cite{Huber:2004ug,Itow:2001ee,Ayres:2004js},
while $\delta_{\textrm{CP}}$ and
the sign of $\delta m^{2}_{31}$
are to be investigated by future long baseline neutrino oscillation
experiments~\cite{Itow:2001ee,Ayres:2004js}.

In this paper, we consider the search for
the leptonic CP-violation angle
by long baseline neutrino oscillation experiments 
using a conventional $\nu_{\mu}$ superbeam.
CP violation can only be observed through the flavor-changing
processes such as $\nu_{\mu} \to \nu_{\mathrm{e}}$%
\footnote{%
Oscillation of anti-neutrinos such as
$\bar{\nu}_{\mu} \to \bar{\nu}_{\mathrm{e}}$
will not be considered in the present analysis.
}.
It is challenging, however,
to extract the value of $\delta_{\textrm{CP}}$ from this oscillation
since
the $\nu_{\textrm{e}}$ appearance probability is suppressed
by the small value of $\sin^{2} 2 \theta_{13}$.
The search for $\delta_{\textrm{CP}}$ is made more difficult by backgrounds
such as $\nu_{\textrm{e}}$-contamination of the incident beam,
which is inseparable from the signal at the detector,
and the neutral-current events that are misidentified as electron events.
Further obstacles to extracting the value of $\delta_{\textrm{CP}}$
are the uncertainties in the other mixing parameters~%
\cite{bib:param-ambiguity,Huber:2002mx,Donini:2004iv}.
Among them, the uncertainty in the value of $\theta_{13}$,
which is bounded only from above, has been widely studied~%
\cite{bib:param-ambiguity,Huber:2002mx,Donini:2004iv}.
The value of $\theta_{13}$ enters into the leading term of the 
$\nu_{\textrm{e}}$ appearance oscillation probability, and the lack 
of knowledge on its value makes it difficult to search for the 
subleading effect of $\delta_{\textrm{CP}}$.
The ambiguity of $| \delta m^{2}_{31} |$ causes a
similar problem:
the value of $| \delta m^{2}_{31} |$ also enters into
the leading term of the appearance probability
and its current experimental uncertainty is large
enough to obscure the subleading 
$\delta_{\textrm{CP}}$-dependent effect.
This uncertainty can also obscure the dependence on the sign of 
$\delta m^{2}_{31}$ and make the $\delta_{\textrm{CP}}$ search more
difficult.
It is hence necessary to constrain the value of
$| \delta m^{2}_{31} |$ in searching for the CP-violating angle.

The value of $| \delta m^{2}_{31} |$ can be precisely determined
using $\nu_{\mu}$ survival events
accumulated concurrently with the $\nu_{\textrm{e}}$
events~\cite{Huber:2004ug,Huber:2002mx,Donini:2004iv}.
This is due to the large number of $\nu_{\mu}$ events
expected from the large $\nu_{\mu}$-flux available in $\nu_{\mu}$
superbeams.
A combined analysis employing $\nu_{\mu}$ survival events as well as
$\nu_{\textrm{e}}$ appearance events can thus place strong constraints
on the values of $\delta_{\textrm{CP}}$ and $\delta m^{2}_{31}$.
We carry out an example analysis for a test setup
fixing the parameters other than $\delta_{\textrm{CP}}$ and
$\delta m^{2}_{31}$.
We demonstrate that the analysis of the $\nu_{\textrm{e}}$ appearance
events alone does not sufficiently constrain the value of
$\delta_{\textrm{CP}}$ in the presence of the ambiguity of
$| \delta m^{2}_{31} |$
and that the combined analysis efficiently improves the constraint.

This paper is organized as follows.
In Sec.~\ref{sec:osc-prob}, we use an analytic expression of the
$\nu_{\textrm{e}}$ appearance probability to show that the ambiguity
in the leading term can obscure the dependence on
$\delta_{\textrm{CP}}$ and on the sign of $\delta m^{2}_{31}$.
We point out that precise values of $| \delta m^{2}_{31} |$ and
$\sin^{2} 2 \theta_{13} \sin^{2} \theta_{23}$ are necessary to search 
for $\delta_{\textrm{CP}}$.
We then use an analytic expression for the $\nu_{\mu}$ survival
probability to show that it can constrain the value of
$\delta m^{2}_{31}$.
In Sec.~\ref{sec:example-study},
we consider a test setup and calculate allowed regions of parameters
in the $\delta_{\textrm{CP}}$-$\delta m^{2}_{31}$ plane
fixing other parameters.
We show that the combined analysis of
$\nu_{\textrm{e}}$ appearance and $\nu_{\mu}$ survival events,
in principle,
gives a strong constraint
on the value of $\delta_{\mathrm{CP}}$ and on the sign of $\delta m^{2}_{31}$. 
We conclude our work and give discussions in Sec.~\ref{sec:conclusion}.  

%
%
\section{Relevance of the Ambiguity of $\delta m^{2}_{31}$
in the Leptonic CP Violation Search}
\label{sec:osc-prob}
Assuming that the number of neutrino generations is three,
we define
the mixing angles $\theta_{ij}$ ($\{i, j\} \subset \{1, 2, 3\}$),
the CP-violating angle $\delta_{\textrm{CP}}$,
and
the quadratic mass differences $\delta m^{2}_{ij}$
as in Ref.~\cite{Eidelman:2004wy}.
We assume the parameter values to be
$\delta m^{2}_{21} \simeq 8 \times 10^{-5} \mathrm{eV}^{2}$,
$|\delta m^{2}_{31}| \simeq (2\textrm{--}3) \times 10^{-3} \mathrm{eV}^{2}$,
$\sin^{2} \theta_{12} \simeq 0.3$,
$\sin^{2} \theta_{23} \simeq 0.5$, and
$\sin^{2} 2 \theta_{13} \lesssim (0.1\textrm{--}0.2)$~%
\cite{Apollonio:2002gd,Fogli:2005cq},
which accomodate all neutrino oscillation experiments except for
the LSND experiment~\cite{Aguilar:2001ty}.

We consider the case where the neutrino energy $E$ and the baseline
length $L$ satisfies $(\delta m^{2}_{31} L) / (2 E) = O(1)$ so that
the oscillation signal can be observed.
We calculate the $\nu_{\mu} \to \nu_{\textrm{e}}$ oscillation
probability in matter whose density $\rho$, and hence the electron
number density $n_{\textrm{e}}$, is constant.  Taking up to first
order in $\delta m^{2}_{21}$ and in
$
a \equiv
2 \sqrt{2} G_{\mathrm{F}} n_{\mathrm{e}} E =
(7.63 \times 10^{-5} \mathrm{eV}^{2})
(\rho / [\mathrm{g \, cm^{-3}}])
(E / [\mathrm{GeV}]) 
$,
we obtain~\cite{Arafune:1997hd}
\begin{widetext}
\begin{equation}
\begin{split}
  P(\nu_{\mu} \to \nu_{\textrm{e}})
  &
  =
  4 c_{13}^{2} s_{13}^{2} s_{23}^{2}
  \biggl[
  1
  \pm \frac{2 a}{ | \delta m^{2}_{31} | } (1 - 2 s_{13}^{2})
  - \frac{\delta m^{2}_{21} L}{2 E}
    \frac{c_{23} c_{12} s_{12}}{s_{13} s_{23} } \sin \delta_{\textrm{CP}}
  \biggr]
  \\ & \quad \times
  \sin^{2}
  \Biggl[
  \frac{| \delta m^{2}_{31} | L}{4 E}
  \mp \frac{a L}{4 E} (1 - 2 s_{13}^{2})
  \mp \frac{\delta m^{2}_{21} L}{4 E}
  \biggl(
  s_{12}^{2} - \frac{c_{23} c_{12} s_{12}}{s_{13} s_{23} } \cos \delta_{\textrm{CP}}
  \biggr)
  \Biggr],
\end{split}
\label{eq:m2e-order1}
\end{equation}
where $s_{ij}$ and $c_{ij}$ denotes
$\sin \theta_{ij}$ and $\cos \theta_{ij}$, respectively,
and
the top and the bottom of the double sign are taken
when $\delta m^{2}_{31} > 0$ and $\delta m^{2}_{31} < 0$, respectively.
The probability has an oscillatory dependence on the energy
due to the sinusoidal factor, 
while the preceding factor in the first pair of brackets,
depending weakly on the energy, gives the envelope of the oscillation. 
The leading terms of the argument and the envelope
of the sinusoidal factor are given by
$| \delta m^{2}_{31} | L / (4 E)$ and
$4 c_{13}^{2} s_{13}^{2} s_{23}^{2}$, respectively.
The subleading terms of both the argument and the envelope 
depend on $\delta_{\textrm{CP}}$ and the sign of $\delta m^{2}_{31}$.
One can obtain the knowledge on the value of $\delta_{\textrm{CP}}$
and the sign of $\delta m^{2}_{31}$ through subleading effects if
the leading terms are known with sufficient accuracy; if not, the
effects will be obscured by the ambiguity in the leading terms.

The argument of the sinusoidal factor is determined
from the energy which gives the peak of the oscillation probability,
or the peak energy in short.
We thus evaluate the peak energy,
including the corrections from energy dependence of the envelope,
up to first order in $\delta m^{2}_{21}$ and in $a$ to obtain
\begin{equation}
\begin{split}
  E_{\textrm{peak}, n} =
  &
  \frac{1}{(2n + 1) \pi}
  \frac{L}{2}
  \Biggl\{
    |\delta m^{2}_{31}|
    \mp
    \Biggl[
      \biggl( 1 - \frac{4}{[(2n + 1) \pi]^{2}} \biggr)
      \frac{|\delta m^{2}_{31}|}{(2n + 1) \pi}
      a' L (1 - 2 s_{13}^{2})
      + \delta m^{2}_{21} s_{12}^{2}
    \Biggr]
  \\ &
    \pm
    \delta m^{2}_{21}
    \frac{c_{23} c_{12} s_{12}}{s_{13} s_{23} }
    \biggl(
      \cos \delta_{\textrm{CP}} \pm \frac{2}{(2n + 1) \pi} \sin \delta_{\textrm{CP}}
    \biggr)
  \Biggl\},
\end{split}
\label{eq:Epeak-n}
\end{equation}
where $n = 0, 1, 2, \cdots$,
$a' \equiv \sqrt{2} G_{\textrm{F}} n_{\textrm{e}} = a / (2 E) =
(1.93 \times 10^{-4}~\mathrm{km}^{-1}) (\rho / [\mathrm{g~cm^{-3}}])$,
and the top of the double sign is for $\delta m^{2}_{31} > 0$
and the bottom for $\delta m^{2}_{31} < 0$.
Inverting the sign of $\delta m^{2}_{31}$ changes
the signs of the subleading second and third terms
in the braces of Eq.~(\ref{eq:Epeak-n}),
and the dependence on $\delta_{\textrm{CP}}$ appears in the third term.
The ambiguity of the leading term, which we denote by
$\Delta(|\delta m^{2}_{31}|)$,
must be smaller than these subleading terms
to determine $\delta_{\textrm{CP}}$ and the sign of $\delta m^{2}_{31}$
from the observation of $E_{\textrm{peak}, n}$.
The conditions on $\Delta( | \delta m^{2}_{31} | )$ are given, for
typical values of the parameters and $n = 0$, by
\begin{align}
  \Delta( | \delta m^{2}_{31} | )
  & <
  \biggl( 1 - \frac{4}{\pi^{2}} \biggr)
  \frac{|\delta m^{2}_{31}|}{\pi}
  a' L (1 - 2 s_{13}^{2})
  + \delta m^{2}_{21} s_{12}^{2}
  \simeq
  \biggl(
  2 \times 10^{-4}
  \frac{L}{[1000~\mathrm{km}]}
  \frac{\rho}{[2.6~\mathrm{g/cm^{3}}]}
  + 2 \times 10^{-5}
  \biggr)
  \mathrm{eV}^{2},
  \\
  \Delta( |\delta m^{2}_{31}| )
  & <
  \delta m^{2}_{21}
  \frac{c_{23} c_{12} s_{12}}{s_{13} s_{23} }
  \simeq
  \begin{cases}
  2 \times 10^{-4}~\mathrm{eV}^{2}
  \quad \textrm{for $\sin^{2} 2 \theta_{13} = 0.1$}
  \\
  8 \times 10^{-4}~\mathrm{eV}^{2}
  \quad \textrm{for $\sin^{2} 2 \theta_{13} = 0.01$}
  \end{cases}.
\end{align}
The current experimental uncertainty in $|\delta m^{2}_{31}|$
is about $6 \times 10^{-4}~\textrm{eV}^{2}$~\cite{bib:atmospheric},
which is larger or similar to the above critical values.
It is therefore necessary
to constrain the value of $\delta m^{2}_{31}$ within a smaller interval.

Similar analysis of the envelope leads to another set of conditions.
To determine the sign of $\delta m^{2}_{31}$,
we must impose the following condition on the ambiguity of
$4 c_{13}^{2} s_{13}^{2} s_{23}^{2}
 = \sin^{2} 2 \theta_{13} \sin^{2} \theta_{23}$:
\begin{equation}
    \frac{
      \Delta (\sin^{2} 2 \theta_{13} \sin^{2} \theta_{23}) 
    }{
      \sin^{2} 2 \theta_{13} \sin^{2} \theta_{23}
    }
    <
    \frac{2 a}{| \delta m^{2}_{31} |} (1 - 2 s_{13}^{2})
    \simeq
    0.3
    \frac{ L }{[1000~\textrm{km}]}
    \frac{ \rho }{[2.6~\mathrm{g/cm^{3}}]},
    \label{eq:envelope-error-eval-1}
\end{equation}
which is evaluated at $E = E_{\textrm{peak}, 0}$ 
using typical values of
the parameters.
To constrain the value of $\delta_{\textrm{CP}}$,
the condition is
\begin{equation}
    \frac{
      \Delta (\sin^{2} 2 \theta_{13} \sin^{2} \theta_{23}) 
    }{
      \sin^{2} 2 \theta_{13} \sin^{2} \theta_{23}
    }
    <
    \pi
    \frac{ \delta m^{2}_{21} }{ | \delta m^{2}_{31} | }
    \frac{ c_{23} c_{12} s_{12} }{ s_{13} s_{23} }
    \simeq
    \begin{cases}
    0.3 \quad \textrm{for $\sin^{2} 2 \theta_{13} = 0.1$} \\
    0.9 \quad \textrm{for $\sin^{2} 2 \theta_{13} = 0.01$}
    \end{cases}.
    \label{eq:envelope-error-eval-2}
\end{equation}
The current experimental bound on the value of $\theta_{23}$ is
$\sin^{2} 2 \theta_{23} > 0.9$~\cite{bib:atmospheric,Fogli:2005cq}.
The value of $\theta_{13}$, on the other hand, is bound only from 
above to date.
Ignorance of the value of $\theta_{13}$ is shown to be
a major obstacle in determining
the value of $\delta_{\textrm{CP}}$ through long baseline
experiments~\cite{bib:param-ambiguity,Huber:2002mx,Donini:2004iv}.
A possible approach to this difficulty is to combine results
from reactor neutrino experiments~%
\cite{Anderson:2004pk,Huber:2004ug,bib:reactor-minakata,Huber:2003pm}.
Future experiments searching for the disappearance of
$\bar{\nu}_{\textrm{e}}$ from reactors are expected to constrain the 
value of $\theta_{13}$ independently of the values of other parameters 
such as $\delta_{\textrm{CP}}$ and matter density.
Reference~\cite{Huber:2003pm} suggests that future experiments can
constrain the value of $\sin^{2} 2 \theta_{13}$ with $\lesssim 10\%$
accuracy if the value of $\sin^{2} 2 \theta_{13}$ is as large as
$0.1$.
Such reactor experiments can be performed in advance or
concurrently with accelerator-based $\delta_{\textrm{CP}}$ searches.
In this prospect,
we assume that the value of
$\sin^{2} 2 \theta_{13} \sin^{2} \theta_{23}$ will be known
with a reasonable accuracy
by the time of $\delta_{\textrm{CP}}$ searches, and
we keep the value of $\sin^{2} 2 \theta_{13} \sin^{2} \theta_{23}$ 
fixed in this paper
to explore the best-case scenario for the long baseline experiments.
%

%
%
The $\nu_{\mu}$ survival events,
which can be accumulated
simultaneously with the $\nu_{\textrm{e}}$ appearance events,
can be used
to constrain the value of $\delta m^{2}_{31}$
owing to the large statistics available.
The energy dependence of the $\nu_{\mu}$ survival probability is calculated,
up to first order in $\delta m^{2}_{21}$ and in $a$,
as~\cite{Arafune:1997hd}
\begin{equation}
\begin{split}
  P(\nu_{\mu} \to \nu_{\mu})
  & =
  1
  -
  4 c_{13}^{2} s_{23}^{2} (1 - c_{13}^{2} s_{23}^{2})
  \biggl[
  1
  \mp 2 \frac{a}{| \delta m^{2}_{31} |}
    \frac{s_{13}^{2}(1 - 2 c_{13}^{2} s_{23}^{2})}{1 - c_{13}^{2} s_{23}^{2}}
  \biggr]
  \\ & \quad \times
  \sin^{2}
  \bigg[
 \frac{|\delta m^{2}_{31}| L}{4 E}
  \pm
  \frac{a L}{4 E}
  \frac{ s_{13}^{2} (1 - 2 c_{13}^{2} s_{23}^{2})}{ 1 - c_{13}^{2} s_{23}^{2} }
  \mp
  \frac{\delta m^{2}_{21} L}{4 E}
  \frac{ s_{13}^{2} s_{23}^{2} s_{12}^{2}
    + c_{23}^{2} c_{12}^{2}
    - 2 c_{23} s_{23} c_{12} s_{12} s_{13} \cos \delta_{\textrm{CP}}
  }{ 1 - c_{13}^{2} s_{23}^{2} }
  \bigg],
\end{split}
\label{eq:m2m-order1}
\end{equation}
where
the top and the bottom of the double sign in Eq.~(\ref{eq:m2m-order1})
are taken when $\delta m^{2}_{31} > 0$ and $\delta m^{2}_{31} < 0$,
respectively.
Since the value of 
$a s_{13}^{2} (1 - 2 c_{13}^{2} s_{23}^{2})/( 1 - c_{13}^{2} s_{23}^{2})$ 
is negligibly small compared to the leading terms
under the current experimental limits,
the observation of the energy dependence of this mode
gives the value of
\begin{equation}
  \biggl|
    \delta m^{2}_{31} -
    \delta m^{2}_{21}
    \frac{
      s_{13}^{2} s_{23}^{2} s_{12}^{2}
      + c_{23}^{2} c_{12}^{2}
      - 2 c_{23} s_{23} c_{12} s_{12} s_{13} \cos \delta_{\textrm{CP}}
    }{ 1 - c_{13}^{2} s_{23}^{2} }
  \biggr|.
\label{eq:m2m-phase}
\end{equation}
\end{widetext}
We then obtain two possible values of $\delta m_{31}^{2}$,
one being positive and the other being negative; 
their absolute values differ from each other
by twice the $\delta m_{21}^{2}$ factor in Eq.~(\ref{eq:m2m-phase}).
This constraint on $\delta m^{2}_{31}$
would contribute to determining
the value of $\delta_{\textrm{CP}}$
and the sign of $\delta m^{2}_{31}$.
Consequently, a combined analysis of the
$\nu_{\textrm{e}}$ appearance and $\nu_{\mu}$ survival events
is crucial for measuring $\delta_{\textrm{CP}}$.
It gives the simultaneous constraint on
$\delta_{\textrm{CP}}$ and $\delta m^{2}_{31}$,
and thus
the result shall be presented
in the $\delta_{\textrm{CP}}$-$\delta m^{2}_{31}$ plane.
%

%
%
\section{Numerical Analysis of an Example Setup}
\label{sec:example-study}
In this section,
we consider a test experiment setup
and obtain constraints on $\delta_{\textrm{CP}}$ and $\delta m^{2}_{31}$
from the oscillation event spectra,
which are numerically calculated
without employing any approximation formulae
of the oscillation probabilities.
We fix other parameters including $\theta_{13}$ and $\theta_{23}$ as
mentioned in Sec.~\ref{sec:osc-prob}.
We show that the values of $\delta_{\textrm{CP}}$ and $\delta m^{2}_{31}$
are significantly constrained by performing
the combined analysis
of $\nu_{\textrm{e}}$ appearance and $\nu_{\mu}$ survival events.
%

\subsection{Example setup and the method of analysis}
\label{subsec:methods}

\begin{figure}
\begin{center}
  \includegraphics[scale=0.32]{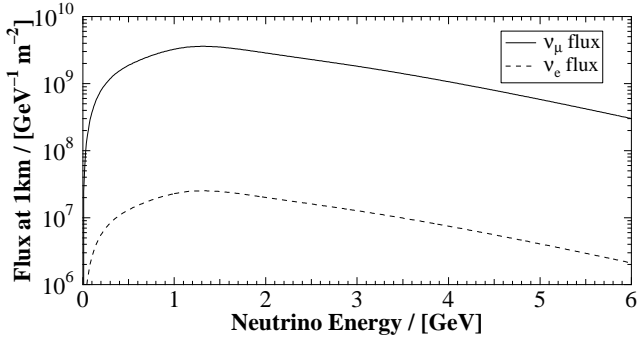}
  \caption{%
  The neutrino fluxes used in the example analysis.
  The flux of muon neutrinos (solid line) is obtained
  by fitting the flux presented in Fig.~3 of Ref.~\cite{Diwan:2003bp}.
  The flux of electron neutrinos (dotted line) is assumed
  to be 0.7\% of that of muon neutrinos with the same
  energy dependence.
  }
  \label{fig:nu-flux}
\end{center}
\end{figure}

%
We consider the following example setup.
A wide band beam of neutrinos
is produced at the upgraded Alternating Gradient
Synchrotron (AGS) at Brookhaven National Laboratory (BNL).
We assume the flux of neutrinos given in Fig.~\ref{fig:nu-flux},
which is obtained by fitting the flux
presented in Fig.~3 of Ref.~\cite{Diwan:2003bp}.
Neutrinos are detected by a water \v{C}erenkov detector
which has 500~kt of fiducial mass
and is placed 770~km away from BNL
in the Kimballton mine, Virginia~\cite{bib:Kimballton}.
We exclusively consider quasi-elastic events
$\nu_{l} + \mathrm{n} \to l^{-} + \mathrm{p}$
as signals.
We assume that
its detection efficiency is unity,
the data acquisition time is $5 \times 10^{7}~\mathrm{sec}$,
and the matter density is $2.6~\mathrm{g / cm^{3}}$.

\begin{figure}
\begin{center}
  \includegraphics[scale=0.32]{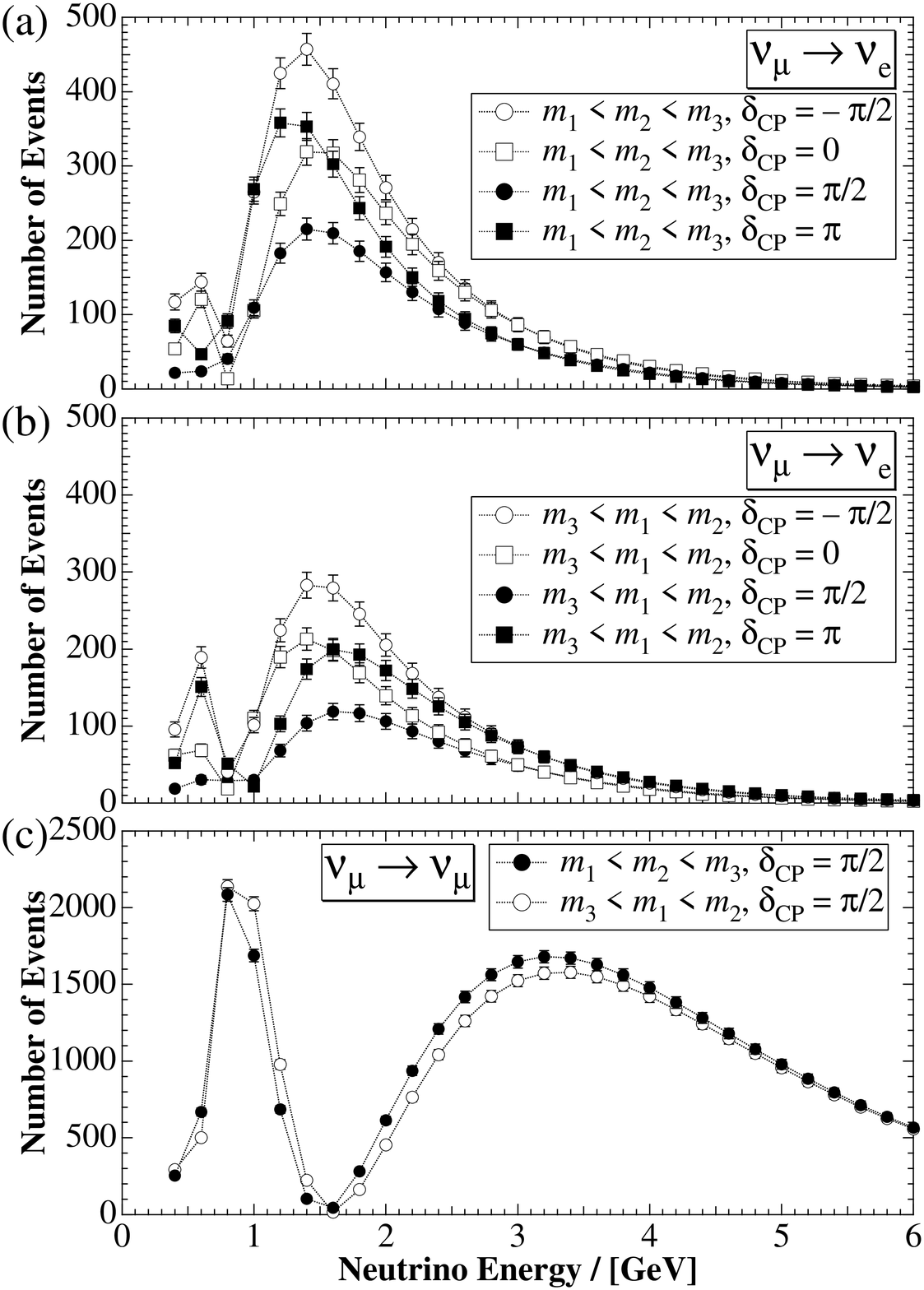}
  \caption{%
  Calculated energy spectra of
  $\nu_{\textrm{e}}$ appearance events (a, b)
  and
  $\nu_{\mu}$ survival events (c)
  for the setup described in text with
  $\delta m^{2}_{31} = \pm2.5 \times 10^{-3}~\mathrm{eV}^{2}$
  ($+$ in (a), $-$ in (b), and both in (c)).
  Other parameters are taken as
  $\delta m^{2}_{21} = 8.2 \times 10^{-5}~\mathrm{eV}^{2}$,
  $\sin^{2} 2 \theta_{12} = 0.84$,
  $\sin^{2} 2 \theta_{23} = 1.00$, and
  $\sin^{2} 2 \theta_{13} = 0.06$.
  The $\nu_{\mu}$ survival event spectra in (c)
  are presented only for $\delta_{\textrm{CP}} = \pi/2$
  since they depend little on $\delta_{\textrm{CP}}$.
  The error bars include only statistical errors.}
  \label{fig:eventnum-spectra}
\end{center}
\end{figure}

%
The expected number of $\nu_{l}$ events,
where $l \in \{ \mathrm{e}, \mu \}$, within an energy bin
$E_{i} < E < E_{i + 1}$ is evaluated as
\begin{equation}
\begin{split}
  &
  \langle N_{i} (\nu_{\mu} \to \nu_{l}) \rangle
  \equiv
  \\ &
  T
  \mathcal{N}
  \int_{E_{i}}^{E_{i + 1}} \mathrm{d} E \,
  \varepsilon(E)
  \frac{ f^{(\nu_{\mu})}(E) }{ L^{2} }
  P(\nu_{\mu} \to \nu_{l})
  \frac{ \mathrm{d} \sigma^{(\nu_{l})} }{ \mathrm{d} E },
\end{split}
  \label{eq:count-in-a-bin}
\end{equation}
where $T$ is the data acquisition time,
$\mathcal{N}$ is the number of target nucleons
in the fiducial volume of the detector,
$\varepsilon(E)$ is the detection efficiency,
$f^{(\nu_{\mu})}(E)$ is the incident $\nu_{\mu}$ flux,
and
$\mathrm{d} \sigma^{(\nu_{l})}  / \mathrm{d} E$ is the
cross section of the detection reaction~\cite{nu-cross-section}.
%
We present in Fig.~\ref{fig:eventnum-spectra}
the calculated event number spectra of
$\nu_{\textrm{e}}$ appearance events and
those of $\nu_{\mu}$ survival events.
The parameters are taken as
$|\delta m^{2}_{31}| = 2.5 \times 10^{-3} \mathrm{eV}^{2}$,
$\delta m^{2}_{21} = 8.2 \times 10^{-5} \mathrm{eV}^{2}$,
$\sin^{2} 2 \theta_{12} = 0.84$,
$\sin^{2} 2 \theta_{23} = 1.00$, and
$\sin^{2} 2 \theta_{13} = 0.06$.
These figures show clear signals of $\nu_{\textrm{e}}$ appearance
and $\nu_{\mu}$ disappearance between 1~GeV and 2~GeV.

%
It is necessary in the analysis to take backgrounds into consideration.
Significant sources of backgrounds
in the $\nu_{\textrm{e}}$ appearance signal
are
$\nu_{\textrm{e}}$-contamination in the incident $\nu_{\mu}$ beam,
and
misidentification of neutral pions
produced through neutral-current interaction.
In the following,
we take into account only the $\nu_{\textrm{e}}$-contamination
which is not separable from signal at the detector.
The number of the neutral-pion background events
is difficult to estimate theoretically
since it depends on the model of the pion-production process,
details of the detector design,
and methods of the event selection;
we leave the consideration of the neutral-pion background to a future work.
The expected number of $\nu_{l}$ events
in an energy bin $E_{i} < E < E_{i + 1}$ is then given by
$\langle N_{i}^{(\nu_{l})} \rangle \equiv
\langle N_{i}(\nu_{\mu} \to \nu_{l}) \rangle
+ \langle N_{i}(\nu_{\textrm{e}} \to \nu_{l}) \rangle$
where $\langle N_{i}(\nu_{\textrm{e}} \to \nu_{l}) \rangle$ is
defined as in Eq.~(\ref{eq:count-in-a-bin})
with the initial $\nu_{\mu}$ replaced by $\nu_{\textrm{e}}$.
We assume in the following that
the flux of $\nu_{\textrm{e}}$-contamination
in the incident beam is 0.7\% of the $\nu_{\mu}$ flux
with the same energy dependence,
as shown in Fig.~\ref{fig:nu-flux} by a dotted line.

%
We obtain the constraint on $\delta_{\textrm{CP}}$ and $\delta m^{2}_{31}$
by the following procedure.
We generate the event number spectrum
$N_{i}^{(\nu_{l})} (\delta_{\textrm{CP}}, \delta m^{2}_{31})$
for various values of $\delta_{\textrm{CP}}$ and $\delta m^{2}_{31}$
while keeping the other parameters fixed.
The number of events in each bin
$N_{i}^{(\nu_{l})} (\delta_{\textrm{CP}}, \delta m^{2}_{31})$
includes statistical errors and
is distributed around
$\langle N_{i}^{(\nu_{l})} (\delta_{\textrm{CP}}, \delta m^{2}_{31}) \rangle$.
We carry out hypothesis testing with the null hypothesis:
``\textit{the parameter values are} $\delta_{\textrm{CP}}^{(\textrm{true})}$
\textit{and} $\delta m^{2 \,(\textrm{true})}_{31}$''.
We reject this null hypothesis if the deviation of
$N_{i}^{(\nu_{l})}(\delta_{\textrm{CP}}, \delta m^{2}_{31})$
from
$\langle
  N_{i}^{(\nu_{l})}
  ( \delta_{\textrm{CP}}^{(\textrm{true})}, \delta m^{2 \, (\textrm{true})}_{31} )
\rangle$
is of statistical significance,
which
we examine by the $\chi^{2}$ test using
\begin{equation}
\begin{split}
  &
  \chi^{2 \, (\nu_{l})}(\delta_{\textrm{CP}}, \delta m^{2}_{31})
  =
  \\ &
  \sum_{i = 1}^{ n_{\textrm{bin}} }
  \frac{
    \bigl|
    N_{i}^{(\nu_{l})}(\delta_{\textrm{CP}}, \delta m^{2}_{31})
    -
    \bigl\langle
      N_{i}^{(\nu_{l})}
      \bigl(
        \delta_{\textrm{CP}}^{(\textrm{true})}, \delta m^{2 \, (\textrm{true})}_{31}
      \bigr)
    \bigl\rangle
    \bigr|^{2}
  }
  { N_{i}^{(\nu_{l})}(\delta_{\textrm{CP}}, \delta m^{2}_{31}) },
\label{eq:chi2-def}
\end{split}
\end{equation}
where $n_{ \textrm{bin} }$ is the number of the energy bins.
The allowed region is obtained as a region in which
the null hypothesis cannot be rejected for a certain confidence level.
%

\subsection{Constraints on the values of $\delta_{\textrm{CP}}$ and $\delta m^{2}_{31}$}
\label{subsec:results}

We carry out the analysis
illustrated in the previous subsection
for three cases:
using $\nu_{\textrm{e}}$ appearance events only,
using $\nu_{\mu}$ survival events only,
and
using both $\nu_{\textrm{e}}$ appearance and
$\nu_{\mu}$ survival events.
In the following,
the values of the parameters
except $\delta m^{2}_{31}$ and $\delta_{\textrm{CP}}$
are fixed to those given in Fig.~\ref{fig:eventnum-spectra},
and we take $|\delta m^{2\,(\textrm{true})}_{31}| = 2.5 \times 10^{-3} \mathrm{eV}^{2}$
and $\delta_{\textrm{CP}}^{(\textrm{true})} = -\pi/2, 0, \pi/2,$ and $\pi$.
%

\begin{figure*}
\begin{center}
  \includegraphics[scale=0.6]{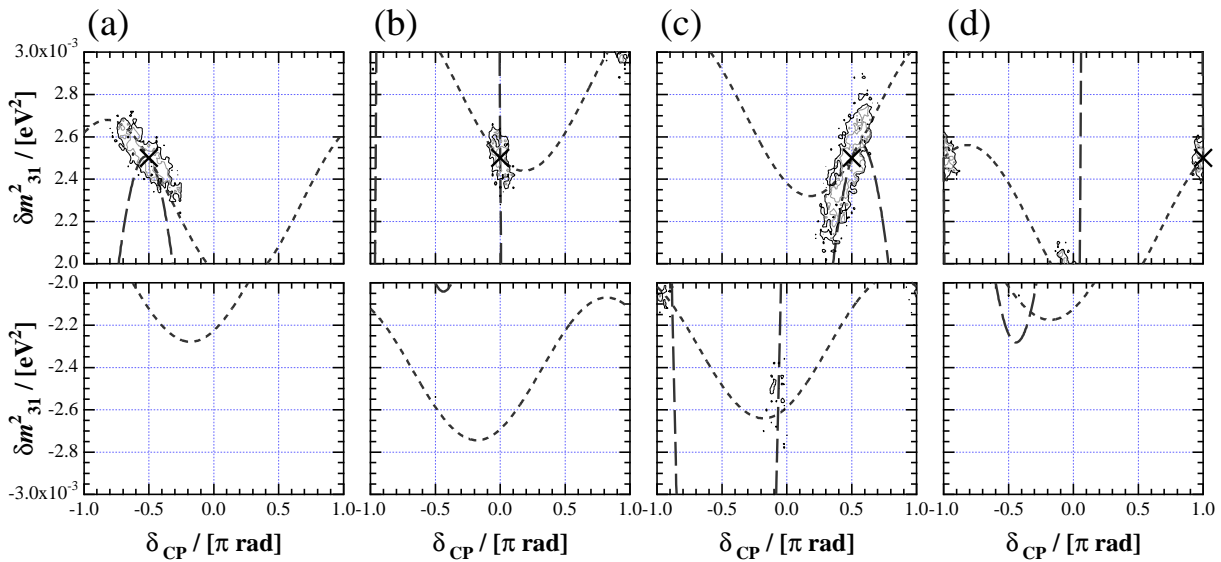}
  \caption{%
  Allowed regions for
  68.3\% (gray solid line) and
  95\% (black solid line) confidence levels
  obtained from $\nu_{\textrm{e}}$ appearance events
  between 0.7~GeV and 3.1~GeV.  The width of energy bins is taken as 0.2~GeV.
  Incident $\nu_{\textrm{e}}$-contamination and statistical errors
  are taken into account.
  The true parameters (cross)
  are taken to be
  $\delta m^{2 \, (\textrm{true})}_{31} = 2.5 \times 10^{-3} \textrm{eV}^{2}$ and
  $\delta_{\textrm{CP}}^{(\textrm{true})} = - \pi/2, 0, \pi/2$, and $\pi$
  in (a), (b), (c), and (d), respectively,
  while $\delta m^{2}_{21}$ and $\sin^{2} 2 \theta_{ij}$'s are
  fixed to those given in Fig.~\ref{fig:eventnum-spectra}.
  The dotted line and the broken line show the solutions to
  Eqs.~(\ref{eq:peakmatch-E}) and (\ref{eq:peakmatch-P}) in text,
  respectively.
  }
  \label{fig:x2e-normal-chi2}
\end{center}
\end{figure*}

\begin{figure*}
\begin{center}
  \includegraphics[scale=0.6]{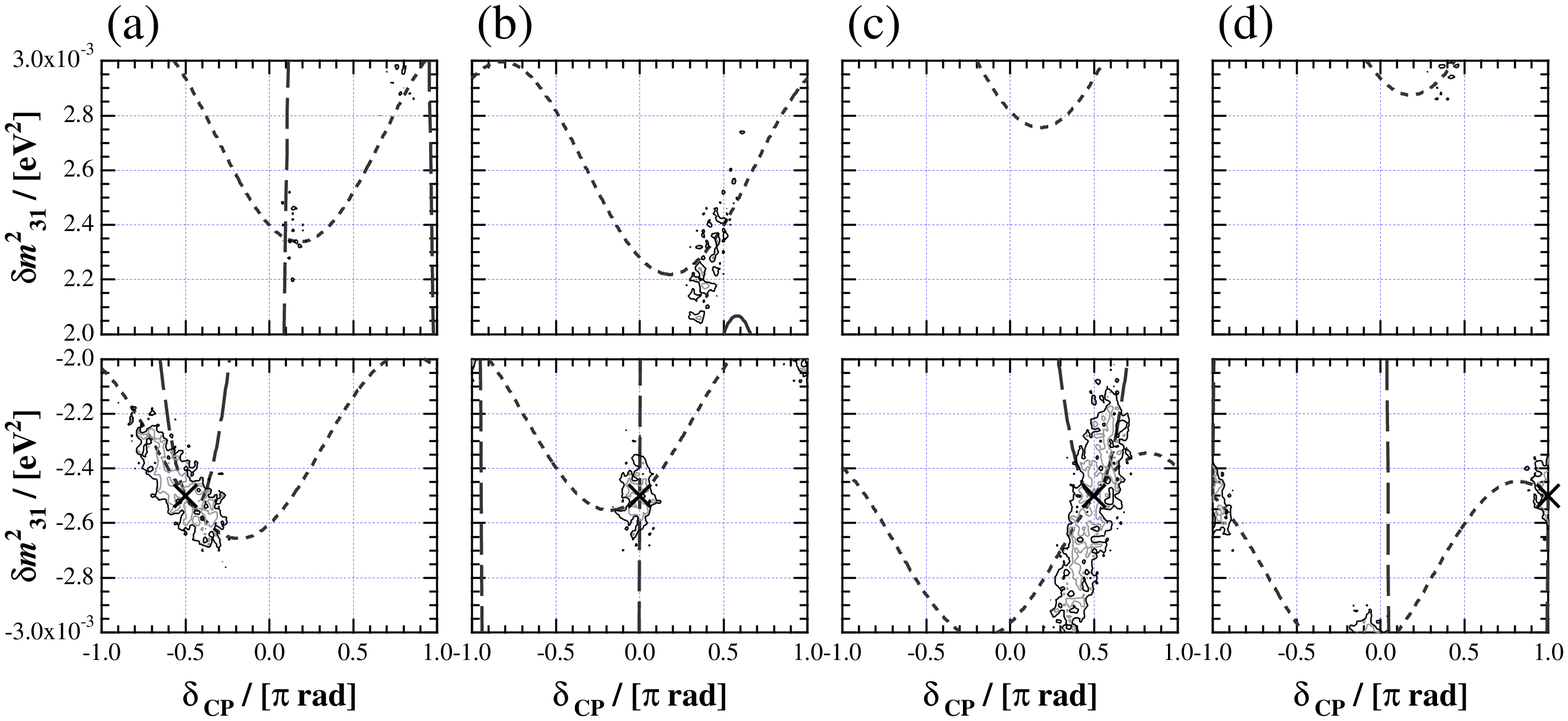}
  \caption{%
  The same as Fig.~\ref{fig:x2e-normal-chi2},
  but
  $\delta m^{2 \, (\textrm{true})}_{31} = - 2.5 \times 10^{-3} \mathrm{eV}^{2}$.
  }
  \label{fig:x2e-inverted-chi2}
\end{center}
\end{figure*}

%
First, we show the analysis of the $\nu_{\textrm{e}}$ appearance spectrum only
using $\chi^{2 \, (\nu_{\textrm{e}})}(\delta_{\textrm{CP}}, \delta m^{2}_{31})$.
We present
in Figs.~\ref{fig:x2e-normal-chi2} and \ref{fig:x2e-inverted-chi2}
allowed regions for 68.3\% and 95\% confidence levels
obtained from the $\nu_{\textrm{e}}$ appearance spectrum.
%
%
We see that $\nu_{\textrm{e}}$ appearance events
cannot always determine the sign of $\delta m^{2}_{31}$
and the value of $\delta_{\textrm{CP}}$.
While there are indeed cases
in which the wrong sign of $\delta m^{2}_{31}$ does not give any allowed
region in the presented figures
(\textit{e.g.} Fig.~\ref{fig:x2e-normal-chi2}(a)),
there are cases where the wrong sign
of $\delta m^{2}_{31}$ also gives allowed regions
(\textit{e.g.} Fig.~\ref{fig:x2e-inverted-chi2}(b)).
The allowed region in such cases, however, is
not a single extended region but a group of small isolated spots,
which indicates that
they are due to statistical errors and background.
%
We also note that
the allowed interval of $\delta_{\textrm{CP}}$
is enlarged by the ambiguity of $\delta m_{31}^{2}$
in some cases such as Fig.~\ref{fig:x2e-normal-chi2} (a);
in this case,
the allowed intervals of the parameters are
$\delta_{\textrm{CP}} \simeq (-0.5 \pm 0.25) \, \pi$
and $\delta m^{2}_{31} \simeq (2.5 \pm 0.2) \times 10^{-3} \, \mathrm{eV}^{2}$
for the 95\% confidence level.
%

%
The allowed regions
in Figs.~\ref{fig:x2e-normal-chi2} and \ref{fig:x2e-inverted-chi2} 
can be understood
in terms of analytic expressions
of Eqs.~(\ref{eq:m2e-order1}) and (\ref{eq:Epeak-n})
as follows.
We expect that the two oscillation spectra resemble each other
and give a small value of $\chi^{2 (\nu_{\textrm{e}})}$
when their peak energies
and peak oscillation probabilities
are equal, \textit{i.e.}
\begin{equation}
  E_{\textrm{peak}} = E_{\textrm{peak}}^{(\textrm{true})},
  \label{eq:peakmatch-E}
\end{equation}
and
\begin{equation}
\begin{split}
  &
  P(\nu_{\mu} \to \nu_{\textrm{e}};
    \delta_{\textrm{CP}},
    \delta m^{2}_{31};
    E_{\textrm{peak}})
  \\ &
  =
  P \bigl(
    \nu_{\mu} \to \nu_{\textrm{e}};
    \delta_{\textrm{CP}}^{(\textrm{true})},
    \delta m^{2 \, (\textrm{true})}_{31};
    E_{\textrm{peak}}^{(\textrm{true})}
  \bigr),
\end{split}
\label{eq:peakmatch-P}
\end{equation}
where
$E_{\textrm{peak}}$ and $E_{\textrm{peak}}^{(\textrm{true})}$
are given by Eq.~(\ref{eq:Epeak-n}) with $n = 0$.
Solutions to these equations are shown
in Figs.~\ref{fig:x2e-normal-chi2} and \ref{fig:x2e-inverted-chi2}
by
a dotted line for Eq.~(\ref{eq:peakmatch-E}) and
a broken line for Eq.~(\ref{eq:peakmatch-P}).
The allowed values of $\delta_{\textrm{CP}}$ and $\delta m^{2}_{31}$
are well constrained around the region where the dotted and the broken lines
intersect or get close to each other,
even for cases with the wrong sign of $\delta m^{2}_{31}$.
It shows that it is difficult to distinguish two spectra
satisfying the peak-matching conditions
Eqs.~(\ref{eq:peakmatch-E}) and (\ref{eq:peakmatch-P}).
Difference of the two spectra away from the peak energy
is not significant under the limited statistics and background.
%

\begin{figure}
\begin{center}
  \includegraphics[scale=0.6]{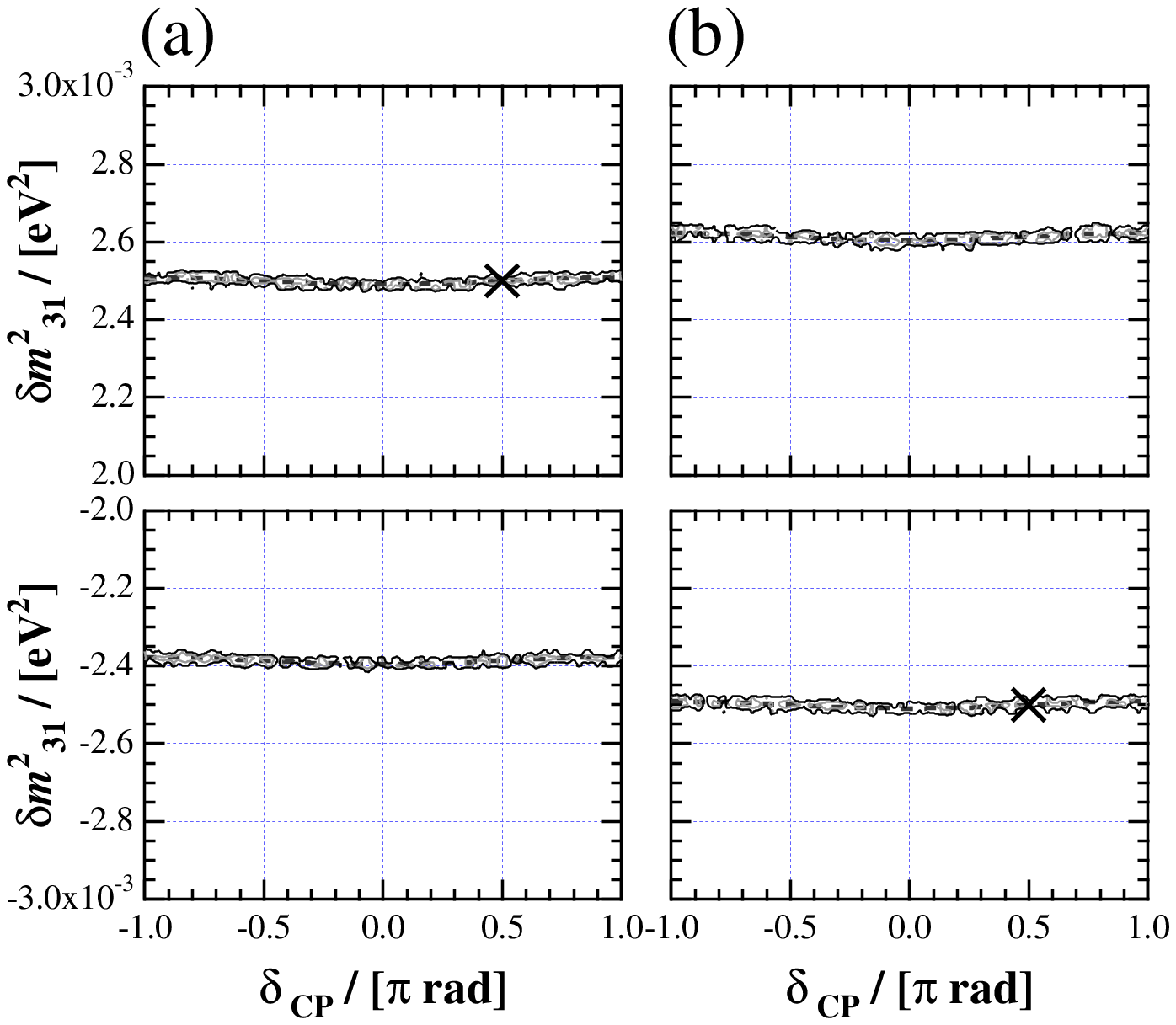}
  \caption{%
  Allowed regions for 68.3\% (gray solid line)
  and 95\% (black solid line) confidence level
  obtained from $\nu_{\mu}$ survival events
  between 0.7~GeV and 4.1~GeV.
  Incident $\nu_{\textrm{e}}$-contamination
  as well as statistical errors are taken into account.
  The true parameters (cross) are
  $\delta_{\textrm{CP}}^{(\textrm{true})} = \pi/2$ and
  (a)
  $\delta m^{2 \, (\textrm{true})}_{31} = 2.5 \times 10^{-3} \mathrm{eV}^{2}$,
  (b)
  $\delta m^{2 \, (\textrm{true})}_{31} = - 2.5 \times 10^{-3} \mathrm{eV}^{2}$,
  while $\delta m^{2}_{21}$ and $\sin^{2} 2 \theta_{ij}$'s are
  fixed to those given in Fig.~\ref{fig:eventnum-spectra}.
  The dotted line is explained in text.
  }
  \label{fig:x2m-chi2}
\end{center}
\end{figure}

%
Second,
we present allowed regions
obtained from $\nu_{\mu}$ survival spectra
in Fig.~\ref{fig:x2m-chi2}.
We show the result
only for $\delta_{\textrm{CP}}^{(\textrm{true})} = \pi/2$
since the results depend little on $\delta_{\textrm{CP}}^{(\textrm{true})}$.
The allowed region separates into two horizontal bands,
one with $\delta m^{2}_{31} > 0$ and the other with $\delta m^{2}_{31} < 0$.
The error of the two values of $\delta m^{2}_{31}$
shrinks down to about $\pm 3 \times 10^{-5}~\mathrm{eV}^{2}$
at the 95\% confidence level.
This small error is due to the large statistics available.
In terms of Eq.~(\ref{eq:m2m-order1}),
the two bands correspond to the two possible values of $\delta m^{2}_{31}$
that give the same value of Eq.~(\ref{eq:m2m-phase})
and, consequently,
the same energy dependence of the $\nu_{\mu}$ survival probability
given in Eq.~(\ref{eq:m2m-order1}).
These two values are indicated in Fig.~\ref{fig:x2m-chi2}
by dotted lines which lie inside the allowed regions.

\begin{figure*}
\begin{center}
  \includegraphics[scale=0.6]{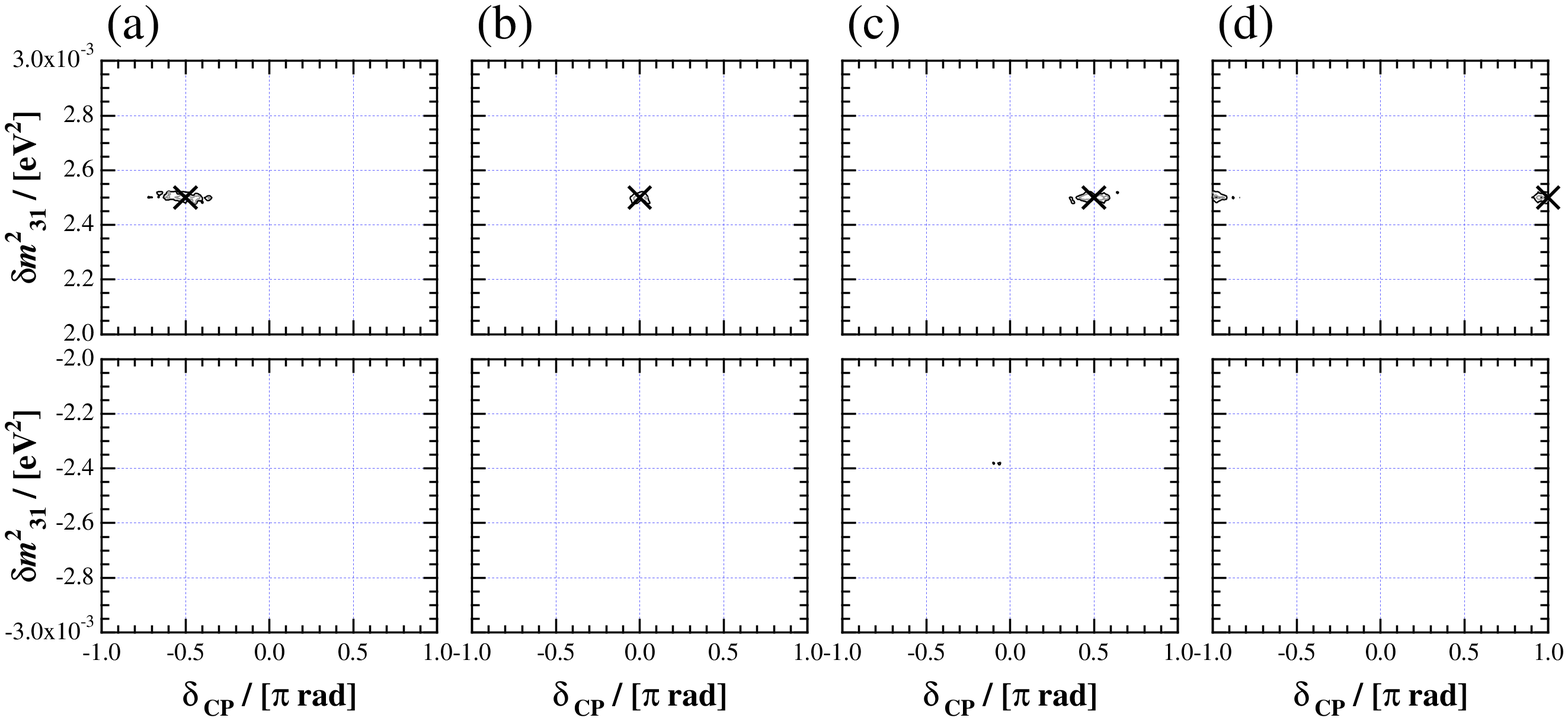}
  \caption{%
  Allowed regions for 68.3\% (gray solid line)
  and 95\% (black solid line) confidence level.
  The values of the parameters are the same as in Fig.~\ref{fig:x2e-normal-chi2}.
  }
  \label{fig:nucombo-normal-chi2}
\end{center}
\end{figure*}

\begin{figure*}
\begin{center}
  \includegraphics[scale=0.6]{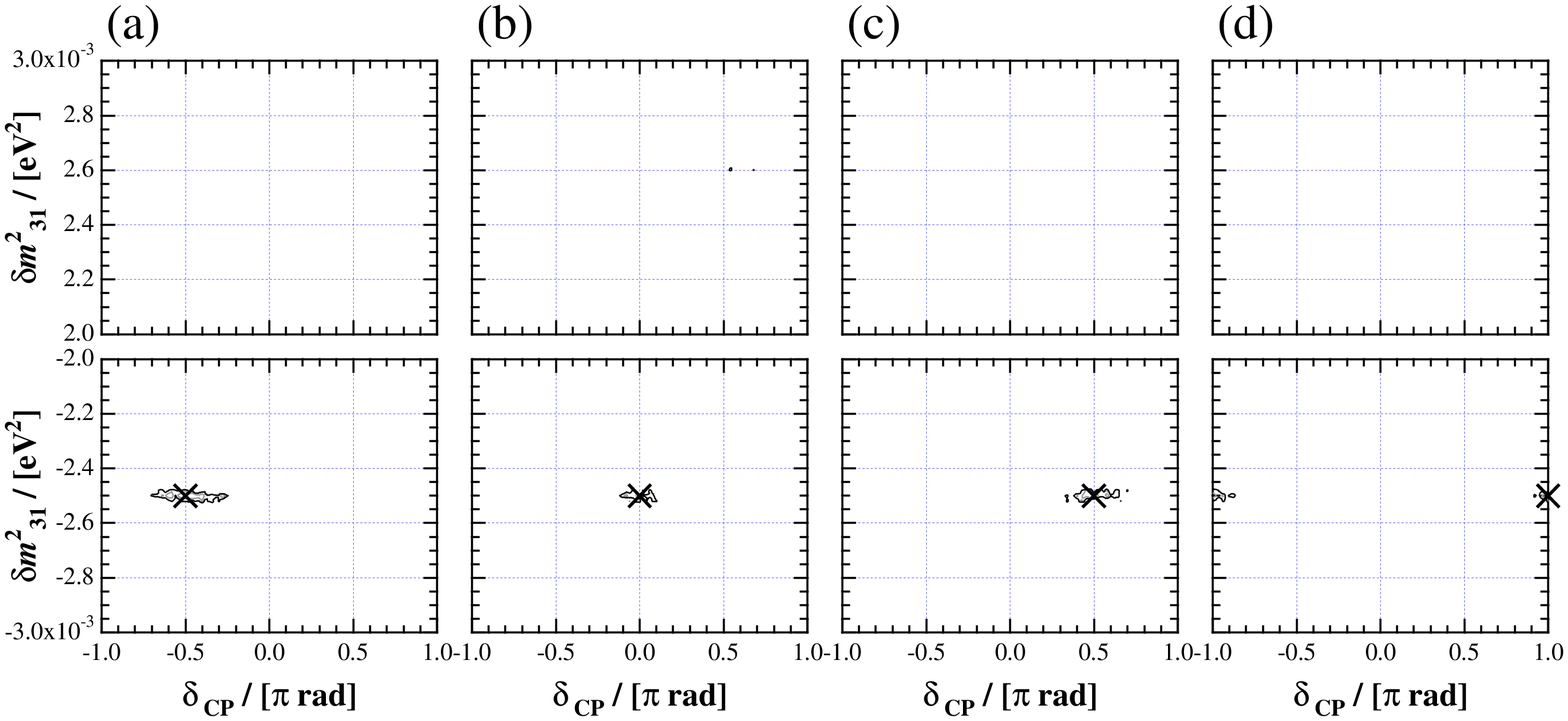}
  \caption{%
  The same as Fig.~\ref{fig:nucombo-normal-chi2},
  but $\delta m^{2 \, (\textrm{true})}_{31} = -2.5 \times 10^{-3} \textrm{eV}^{2}$.
  }
  \label{fig:nucombo-inverted-chi2}
\end{center}
\end{figure*}

%
Finally, we present the combined analysis by evaluating
$\chi^{2} = \chi^{2 \, (\nu_{\textrm{e}})} + \chi^{2 \, (\nu_{\mu})}$.
The allowed regions from the combined analysis are presented in
Figs.~\ref{fig:nucombo-normal-chi2} and \ref{fig:nucombo-inverted-chi2}.
It is shown that
the isolated allowed regions are eliminated in these figures.
The sign of $\delta m^{2}_{31}$ is hence well determined,
and
its absolute value is restricted to
$(2.5 \pm 0.03) \times 10^{-3}~\mathrm{eV}^{2}$,
whose small error also reduces the error of $\delta_{\textrm{CP}}$ down to about
$\pm 0.15~\pi$ or less around the true value.
The synergy of $\nu_{\textrm{e}}$ appearance events
and $\nu_{\mu}$ survival events
excludes the fake allowed regions with the wrong sign of $\delta m^{2}_{31}$
and improves the precision of the parameters.
The observation of both event types and the combined analysis
are thus crucial.

%
%
\section{Conclusion and Discussions}
\label{sec:conclusion}

We considered a search for the CP-violating angle $\delta_{\mathrm{CP}}$ 
in long baseline neutrino oscillation experiments.
We pointed out that
it is necessary
to take 
the ambiguities in $| \delta m^{2}_{31} |$, $\theta_{13}$ and $\theta_{23}$
into consideration
in CP-violation searches.
Their ambiguities can obscure
the dependence of the $\nu_{\mu} \to \nu_{\textrm{e}}$
oscillation probability on $\delta_{\textrm{CP}}$ and
on the sign of $\delta m^{2}_{31}$.
We then showed that
the $\nu_{\mu}$ survival events can be employed
to precisely determine the value of $\delta m^{2}_{31}$,
and be combined with $\nu_{\textrm{e}}$ appearance events
to improve constraints on the value of $\delta_{\textrm{CP}}$
and to determine the sign of $\delta m^{2}_{31}$.
%

%
We numerically verified the significance of the combined analysis
for an example setup.
We assumed that 
the neutrino beam generated by the upgraded AGS beam at BNL
is observed for $5 \times 10^{7}~\mathrm{sec}$
by a 500~kt water \v{C}erenkov detector,
placed 770~km away at Kimballton mine.
We took into account the ambiguity of $| \delta m^{2}_{31} |$
and fixed parameters other than $\delta m^{2}_{31}$ and $\delta_{\textrm{CP}}$.
We considered only the $\nu_{\textrm{e}}$-contamination
in the incident $\nu_{\mu}$ beam as a source of background.
We obtained allowed regions
in the $\delta_{\textrm{CP}}$-$\delta m^{2}_{31}$ plane
through a $\chi^{2}$ analysis.
The analysis of $\nu_{\textrm{e}}$ appearance events alone
led to less restrictive constraints on the value of $\delta_{\textrm{CP}}$
and the sign of $\delta m^{2}_{31}$
due to the ambiguity of $|\delta m^{2}_{31}|$.
In contrast,
the combined analysis 
of $\nu_{\textrm{e}}$ appearance and $\nu_{\mu}$ survival events 
was capable of 
constraining the values of $\delta_{\mathrm{CP}}$ and
$| \delta m^{2}_{31} |$
with the errors of $\pm 0.15~\pi$ or less
and $\pm 3 \times 10^{-5}~\mathrm{eV}^{2}$, respectively,
and also of determining the sign of $\delta m^{2}_{31}$.
%

%
The values of $\theta_{13}$ and $\theta_{23}$,
which we have kept fixed in the present analysis,
may not be known with sufficient precision by the time of
$\delta_{\textrm{CP}}$ searches.
The ambiguity in the value of
$\sin^{2} 2 \theta_{13} \sin^{2} \theta_{23}$
can actually be a hurdle in searching
for the value of $\delta_{\textrm{CP}}$ and the sign of
$\delta m^{2}_{31}$,
as discussed in Sec.~\ref{sec:osc-prob} and
in other studies~%
\cite{bib:param-ambiguity,Huber:2002mx,Donini:2004iv}.
This, however, does not change our major conclusions,
\textit{i.e.}:
the ambiguity in the value of $| \delta m^{2}_{31} |$ is also
a significant obstacle to the $\delta_{\textrm{CP}}$ search,
and that
ambiguity can be diminished by employing the
$\nu_{\mu}$ survival events.
First,
note that the ambiguities
of $\sin^{2} 2 \theta_{13} \sin^{2} \theta_{23}$
and of $| \delta m^{2}_{31} |$
affect two separate features of the $\nu_{\mu} \to \nu_{\textrm{e}}$
oscillation probability,
namely the amplitude and the phase,
as can be seen in Eq.~(\ref{eq:m2e-order1}).
Since they have independent impacts on
the $\delta_{\textrm{CP}}$-search in our approximation,
the ambiguity of $| \delta m^{2}_{31} |$ will still have
significant effects
independently of $\theta_{13}$ and $\theta_{23}$.
Second,
the phase of the $\nu_{\mu}$ survival probability,
Eq.~(\ref{eq:m2m-phase}),
depends little on
$\theta_{13}$ and $\theta_{23}$
for small $\theta_{13}$ below the current upper bound.
Hence
the value of $| \delta m^{2}_{31} |$ will be
tightly constrained from the $\nu_{\mu}$ disappearance probability
regardless of the ambiguity in $\theta_{13}$ and $\theta_{23}$.
Our considerations are also supported by
Ref.~\cite{Donini:2004iv},
although its authors analyzed a counting
experiment using low-energy neutrinos and did not consider an
experiment measuring the energy of neutrinos as we did.

%
Since the ambiguity of $| \delta m^{2}_{31} |$ has been shown
to be controllable,
the ambiguities of $\theta_{23}$ and $\theta_{13}$ are
the remaining obstacles to $\delta_{\textrm{CP}}$ searches.
Of the two,
the value of $\theta_{23}$ 
can be determined from the amplitude 
of the $\nu_{\mu}$ disappearance probability
(\textit{cf}. Eq.~(\ref{eq:m2m-order1}))~%
\cite{Huber:2004ug,Huber:2002mx,Donini:2004iv,Minakata:2004pg}.
This hardly 
interferes
with the determination of $| \delta m^{2}_{31} |$
since the value of $| \delta m^{2}_{31} |$
is determined from the the phase of the oscillation.
The improvement of the uncertainty in $\theta_{13}$
must await future experiments
as discussed in Section \ref{sec:osc-prob}.
%

%
The combined analysis presented in this paper
can be applied to more realistic case studies.
For that purpose, it is necessary to take account of
backgrounds other than $\nu_{\textrm{e}}$-contamination,
especially that from single pion production events.
A neutral pion becomes a source of background
when the two photons from its decay are not separately detected
due to the limited resolution of the detector
and thereby
misidentified as a $\nu_{\textrm{e}}$ appearance signal.
The estimation of this background requires careful treatments.
It is difficult to theoretically estimate
the number of single-$\pi^{0}$-production events
owing to its dependence on how strong-interaction processes are modeled.
The number of misidentified events among them
further depends on the details
of the setup and method of experiments such as
the design of the detector
and the criteria of event selection.
Nevertheless,
there is a case study claiming that
the number of background events from the neutral pions
can be suppressed to be comparable to
that from the incident $\nu_{\textrm{e}}$-contamination~\cite{Itow:2001ee}.
We expect
that the analysis presented in this paper shall remain effective
when such excellent suppression of background is possible.
We leave a realistic analysis
with the consideration of $\pi^{0}$-background for a future work.
%

%
%
\begin{acknowledgments}

The authors acknowledge
Professor Robert Bruce Vogelaar and Professor Raju Raghavan
for helpful discussions.
This research was supported in part by the U.S. Department of Energy, 
grant DE--FG05--92ER40709, Task A (T.T.)
and by the U.S. National Science Foundation, grant PHY--9972127 (M.K.).

\end{acknowledgments}

%
%

\end{document}